\documentclass[letter]{aa}  
\usepackage{graphicx}
\usepackage{amsmath}
\usepackage{amsfonts}
\usepackage{amssymb}
\usepackage{gensymb}
\usepackage{textcomp}
\usepackage[varg]{txfonts}
\usepackage[T1]{fontenc}
\usepackage{natbib}
\usepackage{hyperref}
\hypersetup{
  colorlinks   = true, 
  urlcolor     = blue, 
  linkcolor    = blue, 
  citecolor   = blue 
}

\bibpunct{(}{)}{;}{a}{}{,}
\begin{document}

\title{Hard X-rays from the deep solar atmosphere:}
\subtitle{An unusual UV burst with flare properties}

\author{
L.~P.~Chitta\inst{1}, 
I.~G.~Hannah\inst{2}, 
L. Fletcher\inst{2,3}, 
H.~S.~Hudson\inst{2,4}, 
P.~R.~Young\inst{5,6}, 
S.~Krucker\inst{4,7},
and
H.~Peter\inst{1,8}
}

\institute{
Max-Planck-Institut f\"{u}r Sonnensystemforschung, Justus-von-Liebig-Weg 3, 37077 G\"{o}ttingen, Germany\\
\email{chitta@mps.mpg.de} 
\and
School of Physics \& Astronomy, University of Glasgow, University Avenue, Glasgow, G12 8QQ, UK
\and
Rosseland Centre for Solar Physics, University of Oslo, PO Box 1029 Blindern, NO-0315 Oslo, Norway
\and
Space Sciences Laboratory University of California, Berkeley, CA 94720, USA
\and
NASA Goddard Space Flight Center, Greenbelt, MD 20771, USA
\and
Northumbria University, Newcastle upon Tyne, NE1 8ST, UK
\and
University of Applied Sciences and Arts Northwestern Switzerland, CH-5210 Windisch, Switzerland
\and
Institut f\"{u}r Sonnenphysik (KIS), Georges-K\"{o}hler-Allee 401a, 79110 Freiburg, Germany
}
   \date{Received ; accepted }
   
\abstract
{Explosive transient events occur throughout the solar atmosphere. The differing manifestations range from coronal mass ejections to Ellerman bombs. The former may have negligible signatures in the lower atmosphere, and the latter may have negligible nonthermal emissions such as hard X-radiation. A solar flare generally involves a broad range of emission signatures. Using a suite of four space-borne telescopes, we report a solar event that combines aspects of simple UV bursts and hard X-ray emitting flares at the same time. The event is a compact C-class flare in active region AR11861, SOL2013-10-12T00:30. By fitting a combined isothermal and nonthermal model to the hard X-ray spectrum, we inferred plasma temperatures in excess of 15\,MK and a nonthermal power of about $3\times10^{27}$\,erg\,s$^{-1}$ in this event. Despite these high temperatures and evidence for nonthermal particles, the flare was mostly confined to the chromosphere. However, the event lacked clear signatures of UV spectral lines, such as the Fe\,{\sc xii} 1349\,\AA\ and  Fe\,{\sc xxi} 1354\,\AA\ emission lines, which are characteristic of emission from hotter plasma with a temperature over 1\,MK. Moreover, the event exhibited very limited signatures in the extreme-UV wavelengths. Our study indicates that a UV burst -- hard X-ray flare hybrid phenomenon exists in the low solar atmosphere. Plasma that heats to high temperatures coupled with particle acceleration by magnetic energy that is released directly in the lower atmosphere sheds light on the nature of active region core heating and on inferences of flare signatures.}

\titlerunning{Hard X-rays from the deep solar atmosphere: An unusual UV burst with flare properties}
\authorrunning{L.~P.~Chitta et al.}

   \maketitle

\section{Introduction\label{sec:int}}

The atmospheric layers in the solar chromosphere and transition region (TR) harbor discrete ultraviolet (UV) brightening events in the quiet Sun and in active regions. These brightenings, also called explosive events, in the quiet Sun were first discovered with data from the Naval Research Laboratory's High Resolution Telescope and Spectrograph \citep[][]{1983ApJ...272..329B,1989SoPh..123...41D}. They are generally considered to be triggered by magnetic reconnection in the solar atmosphere \citep[][]{1989SoPh..123...41D,1997Natur.386..811I}. UV bursts in active regions are compact 1\arcsec--2\arcsec\ small-scale brightening events that are often observed in emerging flux regions \citep[][]{2014Sci...346C.315P,2018SSRv..214..120Y}. The Interface Region Imaging Spectrograph \citep[IRIS;][]{2014SoPh..289.2733D} has been providing valuable imaging spectroscopic diagnostics to improve our understanding of these small-scale phenomena. Numerical simulations show that the UV burst emission signatures can originate from the low chromosphere when the plasma is explosively heated through magnetic reconnection to temperatures over 0.02\,MK \citep[][]{2022A&A...665A.116N,2024A&A...685A...2C}. A salient characteristic of UV bursts is the superposition of absorption lines (indicative of cooler plasma) on emission lines that generally form at temperatures of 0.02\,MK--0.1\,MK and sample plasma from the chromosphere to the TR \citep{2016A&A...590A.124R}. These emission lines are usually formed in the upper chromosphere to the TR, but in the case of UV bursts, they are overlaid with absorption lines. This implies that cool material lies above the hot material and that the formation height is much lower than usual \citep[][]{2014Sci...346C.315P}.

Inversions of spectral line profiles including nonlocal thermodynamic equilibrium effects suggest that UV bursts might be heated to a few 10,000\,K \citep[][]{2019A&A...627A.101V}. UV bursts have connections to other explosive phenomenon on the Sun, namely Ellerman bombs, which occur at photospheric heights \citep[][]{2016ApJ...824...96T}. However, there is only limited evidence that Ellerman bombs and/or UV bursts impact the corona over 1\,MK \citep[e.g.,][]{2004ApJ...601..530S,2019ApJ...871...82G}. Therefore, UV bursts are considered to be a phenomenon mostly in the lower atmosphere that does not show signatures in the coronal extreme-UV (EUV) emission. Nevertheless, based on rough energy estimates, bright UV bursts possess an energy content of $10^{29}$\,erg \citep[][]{2014Sci...346C.315P}, which is comparable to the energy of a small flare \citep[][]{2008ApJ...677..704H,2011SSRv..159..263H}.
 
On the other hand, flares are the most powerful explosions in the heliosphere. They probably originate in the coronal field and have significant coronal emission. Hard X-ray (HXR) and soft X-ray flux enhancements are commonly associated with flaring activity. The flare signatures span almost the entire electromagnetic spectrum, however, and also imprint their signatures on the lower atmosphere \citep[][]{2011SSRv..159...19F,2017LRSP...14....2B}. 
The Reuven Ramaty High-Energy Solar Spectroscopic Imager \citep[RHESSI;][]{2002SoPh..210....3L} had a significant impact on flare research. Early results from RHESSI include observations providing circumstantial evidence for the formation of a current sheet in a flare \citep[][]{2003ApJ...596L.251S}. RHESSI data also revealed double HXR sources over flaring loops as magnetic reconnection signatures \citep[][]{2004ApJ...612..546S}. In spite of the great spectral breadth of flare emissions, we note that major flare and/or coronal mass ejection events with negligible UV or other lower-atmosphere emissions can indeed occur \citep[e.g.,][]{1996JGR...10113497M}

With its high-resolution view of the Sun, the IRIS instrument has also provided crucial insight into the atmospheric signatures of flares. This includes observations with a similar temporal evolution of upflows of 10\,MK\ hot plasma \citep[][]{2015ApJ...807L..22G}. There is also evidence for fluorescently excited molecular hydrogen emission in the much cooler atmosphere \citep[][]{2021MNRAS.504.2842M}.

Magnetic reconnection is a common thread that connects the explosive energy release observed in UV bursts and flares. These impulsive events follow a power-law distribution in energy from large flares to smaller nanoflares \citep[][]{1991SoPh..133..357H}. Two open questions that concern this line of research are (1) the contribution of impulsive heating events to coronal heating, and (2) whether there is a single way to study flares and UV bursts, which are disparate phenomena in terms of their onset location.

To answer question (2) above, we provide evidence for a hybrid explosive phenomenon on the Sun that combines the salient signatures of UV bursts and flares. In the following, we explore these properties and discuss the implications of this event for coronal heating and for energy transport in flares.

\section{Observations\label{sec:obs}}

We made use of observations from four different spacecrafts: the Solar Dynamics Observatory \citep[SDO;][]{2012SoPh..275....3P}, IRIS, RHESSI, and the Geostationary Operational Environmental Satellites (GOES). 

Generally, the SDO provides multiwavelength coverage of the solar atmosphere, including photospheric magnetic fields. The IRIS observations cover imaging and spectroscopy of solar chromospheric and transition region plasma. RHESSI records images and spectra of HXR above 3\,keV that are produced in energetic events such as solar flares. GOES records disk-integrated soft X-ray flux in the 0.5-4\,\AA\ and 1-8\,\AA\ bands. 

The IRIS slit scanned the trailing negative-polarity sunspot in active region NOAA AR11861 using the 400-step \texttt{Very large dense raster} mode between 2013 October 11 UT\,23:54:49 and 2013 October 12 UT\,03:29:36, with a field of view centered at solar $(x,y)=(-151\arcsec,-279\arcsec)$ (OBS ID: 3880013646). These single-raster observations have a spatial scale of 0.16\arcsec\ along the slit direction and a slit width of 0.35\arcsec\ (at the solar disk center, $1\arcsec\,\approx\,725$\,km). IRIS acquired a readout of the entire spectrum in its near-UV (NUV) and far-UV (FUV) wavelength range. The spectral sampling is about 25.5\,m\AA\,pixel$^{-1}$ in the NUV and about 13\,m\AA\,pixel$^{-1}$ in the FUV. The spectra were recorded with an exposure time of 30\,s per slit position.

\begin{figure*}
\begin{center}
\includegraphics[width=0.8\textwidth]{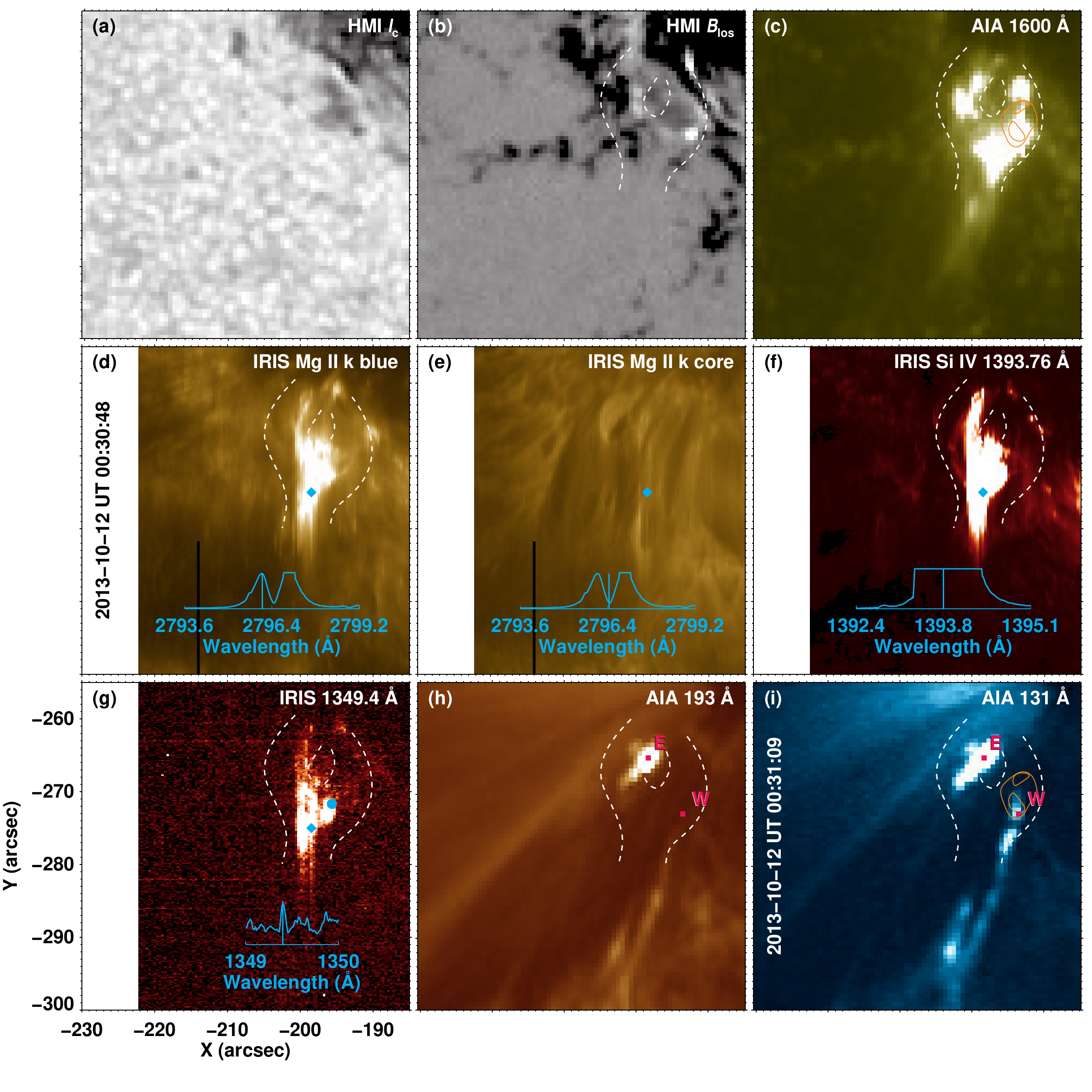}
\caption{Multiwavelength overview of SOL2013-10-12T00:30. Panels (a) and (b) show maps of photospheric continuum intensity and line-of-sight magnetic flux density (saturated at $\pm$250\,G) observed with SDO/HMI. Ultraviolet diagnostics of the event through the chromosphere and transition region are displayed (these include the AIA 1600\,\AA\ map in panel c and raster scans of various spectral lines observed with IRIS in panels d-g). The wavelength position of each raster map in panels (d) to (g) is identified with a vertical cyan line that is overlaid on the spectral line profile. In panels (d)-(g), the spectral line profile is obtained from the pixel centered at the cyan diamond (IRIS pixel coordinates $x=67$; $y=565$). The cyan circle in panel g identifies another spatial location of the brightening (IRIS pixel coordinates $x=75$; $y=584$). The detector images obtained from the centers of the diamond and circles are further plotted in Figs.\,\ref{fig:cdet1} and \ref{fig:cdet2}. The time stamp of IRIS spectral line profiles is noted in panel (d). Panels (h) and (i) are AIA 193\,\AA\ and 131\,\AA\ filter maps covering the event. Its two footpoints, E and W, are marked. The curved white lines in panels (b)-(d) and (f)-(i) outline the cusp-shaped morphology of the event. The contours of RHESSI intensities, integrated between UT\,00:31:00 and UT\,00:31:29, from two energy bands are overlaid in orange in panels (c) and (i) (larger outer contour: 5--8\,keV; smaller inner contours: 11--18\,keV). The contours levels are at 0.3 times the peak intensity of the respective energy band. North is up (see Sects.\,\ref{sec:obs} and \ref{sec:uvb} for details). An animation of panels h and i without contours is available online.\label{fig:map}}
\end{center}
\end{figure*}

The prominent spectral lines in the IRIS raster scan include the Mg\,{\sc ii}\,h and k lines, which sample the chromosphere around 0.01\,MK--0.02\,MK; the C\,{\sc ii} line pair (around 1335\,\AA); Si\,{\sc iv} lines (1394\,\AA\ and 1403\,\AA); and a set of O\,{\sc iv} lines (around 1400\,\AA) that sample the transition region plasma in the range of roughly 0.05\,MK--0.15\,MK. The Fe\,{\sc xii} 1349\,\AA\ line and the Fe\,{\sc xxi} 1354\,\AA\ line in the observed wavelength range would in principle give access to coronal plasma at temperatures of 1.3\,MK and 10\,MK, respectively. We used radiometrically calibrated level-2 data in this study. During these observations, IRIS captured a UV brightening event at the southern penumbral edge of the sunspot, between UT\,00:26 to UT\,00:45 on 2013 October 12. 

We used the photospheric continuum intensity and line-of-sight magnetic field maps recorded by the Helioseismic and Magnetic Imager \citep[HMI;][]{2012SoPh..275..207S} on board the SDO. These data were obtained at a cadence of 45\,s and an image scale of about 0.5\arcsec\,pixel$^{-1}$. The Atmospheric Imaging Assembly \citep[AIA;][]{2012SoPh..275...17L} on board the SDO provided UV and EUV images of the solar atmosphere. In particular, we used level-1 AIA 1600\,\AA\ images in the UV wavelength range. In active regions and plages, the 1600\,\AA\ passband receives contributions from the Si\,{\sc i} continuum and the C\,{\sc iv} 1550\,\AA\ doublet forming at 10$^5$\,K \citep[][]{2019ApJ...870..114S}.

We also used EUV images from the 94\,\AA\ (Fe\,{\sc x}: 6.05; Fe\,{\sc xviii}: 6.85), 131\,\AA\ (Fe\,{\sc viii}: 5.6; Fe\,{\sc xxi}: 7.05),  171\,\AA\ (Fe\,{\sc ix}: 5.85), 193\,\AA\ (Fe\,{\sc xii}: 6.2; Fe\,{\sc xxiv}: 7.25), 211\,\AA\ (Fe\,{\sc xiv}: 6.3), 304\,\AA\ (He\,{\sc ii}: 4.7); and 335\,\AA\ (Fe\,{\sc xvi}: 6.45) filters. The information in parentheses includes dominant ion species that contribute to the EUV emission in the respective filters, along with the logarithm of the peak formation temperature of these ions \citep[][]{2010A&A...521A..21O,2012SoPh..275...41B}. These level-1 AIA UV and EUV data have a cadence of 24\,s and 12\,s, respectively, and both have an image scale of 0.6\arcsec\,pixel$^{-1}$. 

First, we deconvolved the AIA EUV level-1 data with the point spread functions (PSF) of the respective channels using the Python package \texttt{AIAPY}  \citep[][]{Barnes2020,Barnes2021}. This step effectively reduced the effects of artifacts in the images that are caused by filter mesh and PSF diffraction patterns. These SDO AIA and HMI data were all processed using the \texttt{aia\_prep} procedure in the solarsoft library \citep[][]{1998SoPh..182..497F}. The resulting processed data have an image scale of 0.6\arcsec\,pixel$^{-1}$, and solar north is oriented up. We focused on a field of view of $46\arcsec\times46\arcsec$, which was centered on the solar disk at ($-207.6\arcsec$,$-277.2\arcsec$), at a reference time of 2013 October 12 UT\,00:31:08. This covered the IRIS event. 

Active region AR11861 was emerging during the course of two days leading up to the observations. During this period, the Sun was not particularly active, but GOES recorded several C-class flares along with an M-class flare on 2013 October 11 around UT\,07:26. While many of these C-class flares are indeed associated with AR11861, the M-class flare was a behind-the-limb event and was not connected to the AR under investigation.

Based on the IRIS slit-jaw and AIA movies, the UV brightening seems to be an active region jet. These jets are usually associated with B- and C-class GOES X-ray flares and occur around sunspots. Accordingly, at the time of the brightening event caught by IRIS, GOES also detected a minor C-class flare with an X-ray flux of about $10^{-6}$\,W\,m$^{-2}$ in the 1--8\,\AA\ band. RHESSI detected a compact HXR source around solar $(x,y)\approx(-200\arcsec,-270\arcsec)$, cospatial with the IRIS event, around UT\,00:31. We processed the RHESSI imaging data with the maximum entropy method algorithm\footnote{developed at the University of Genoa} and acquired HXR spatial maps in two energy bins (5--8\,keV and 11--18\,keV). To study the temporal evolution of the HXR emission, we also obtained count rates as a function of time in three energy bins (3--6\,keV, 6--12\,keV, and 12--25\,keV from detector 1 of RHESSI). 

To further explore the characteristics of our event, we compared its FUV spectral profiles with data from the Skylab NRL SO82B spectrograph \citep[][]{1977ApOpt..16..879B} of an X-class flare observed in active region NOAA AR209 on 1973 September 7 around UT\,12:12 \citep[][]{1992ApJ...391..393D}. These Skylab data were calibrated by \citet{2019ApJ...870..114S}. Additionally, we compared the UV brightening spectral profiles with those from the ribbon of an X-class flare that IRIS observed 2014 October 25 (see Appendix\,\ref{sec:app} for details).

\section{UV burst characteristics and spatial morphology\label{sec:uvb}}

During the phase of the flux emergence, AR11861 exhibited several compact UV brightening events in its core and moat-flow regions around the sunspots. The southern penumbral portion of the trailing negative-polarity sunspot in this AR was particularly prolific in producing such events. A zoomed-in view of the solar surface covering this penumbral region is presented in Fig.\,\ref{fig:map}a--b. Frequent interactions between the minor positive-polarity magnetic elements and the dominant polarity of the sunspot were observed at this location. Magnetic reconnection caused by these interactions triggered a compact UV brightening event with a cusp-shaped morphology, as seen in the AIA 1600\,\AA\ passband (Fig.\,\ref{fig:map}c). A brighter fan or dome-like structure along with a fainter spine emission are noticeable. This feature is mainly due to the C\,{\sc iv} 1550\,\AA\ doublet and suggests plasma temperatures of at least 10$^5$\,K (under the assumption of ionization equilibrium).

The Mg\,{\sc ii}\,k spectral profiles in the NUV that emerge from the UV brightening as obtained by IRIS are intriguing. The event is well visible in the blue and red wings of the Mg\,{\sc ii}\,k line profile. The morphology of the blue-wing map, at a Doppler shift of $-34$\,km\,s$^{-1}$ from the nominal line core (displayed in Fig.\,\ref{fig:map}d), is very similar to that of the AIA 1600\,\AA\ map. Owing to the high intensity of the event, the spectral profiles in the red wing of the Mg\,{\sc ii}\,k line are saturated (giving rise to flat spectral tops; see the profile overlaid in Fig.\,\ref{fig:map}d). The Mg\,{\sc ii}\,k line core intensity is much lower than that of the wing, and it mainly shows fibril-like elongated structures. These fibrils form the penumbral canopy of the sunspot and may generally trace the magnetic field direction at this location \citep[][]{2011A&A...527L...8D}. Intriguingly, however, there is no clear trace of the main body of the event (i.e., the bright dome structure) in the core of the Mg\,{\sc ii}\,k line profiles. 

In the IRIS FUV wavelength range, the prominent C\,{\sc ii} and Si\,{\sc iv} spectral lines both show clear signatures of the UV brightening. The Si\,{\sc iv} map at 1394\,\AA\ (Fig.\,\ref{fig:map}f) has an almost similar spatial morphology as is seen in the Mg\,{\sc ii}\,k blue-wing map. Under ionization equilibrium conditions, the Si\,{\sc iv} line is formed at temperatures of about 0.08\,MK. In UV bursts, however, the line may already form at temperatures as low as 15\,kK \citep[][]{2016A&A...590A.124R}. This suggests that the hotter UV burst material underlies the cooler chromospheric canopy. A sample Si\,{\sc iv} line profile from the cusp location also reveals an asymmetry between its blue and red wings. In particular, the blue wing has a sharp vertical feature at around 1393.33\,\AA, whereas the red wing shows the expected drop-off. The vertical feature on the blue-wing side is caused by the Ni\,{\sc ii} absorption feature at around 1393.33\,\AA. This further implies the presence of cooler plasma above the UV brightening, which is also consistent with the existence of Mg\,{\sc ii}\,k core fibrils.

We further examined whether hotter plasma could be spotted in IRIS observations of the Fe\,{\sc xii} line (formation temperature of 1.3\,MK), which is expected to emit around 1349.4\,\AA\ in the FUV. Around this wavelength point, we noticed emission features from the brightening with a nearly identical spatial morphology as the Mg\,{\sc ii}\,k blue wing and Si\,{\sc iv} (Fig.\,\ref{fig:map}g). However, the thermal width of this spectral feature around 1349.4\,\AA\ is smaller than the $\sim$20\,km\,s$^{-1}$ width expected from the Fe\,{\sc xii} line. Therefore, this feature in Fig.\,\ref{fig:map}g is probably not caused by the Fe\,{\sc xii} emission in this wavelength range. We discuss this point further in Sect.\,\ref{sec:spec}. 

\begin{figure*}
\begin{center}
\includegraphics[width=0.65\textwidth]{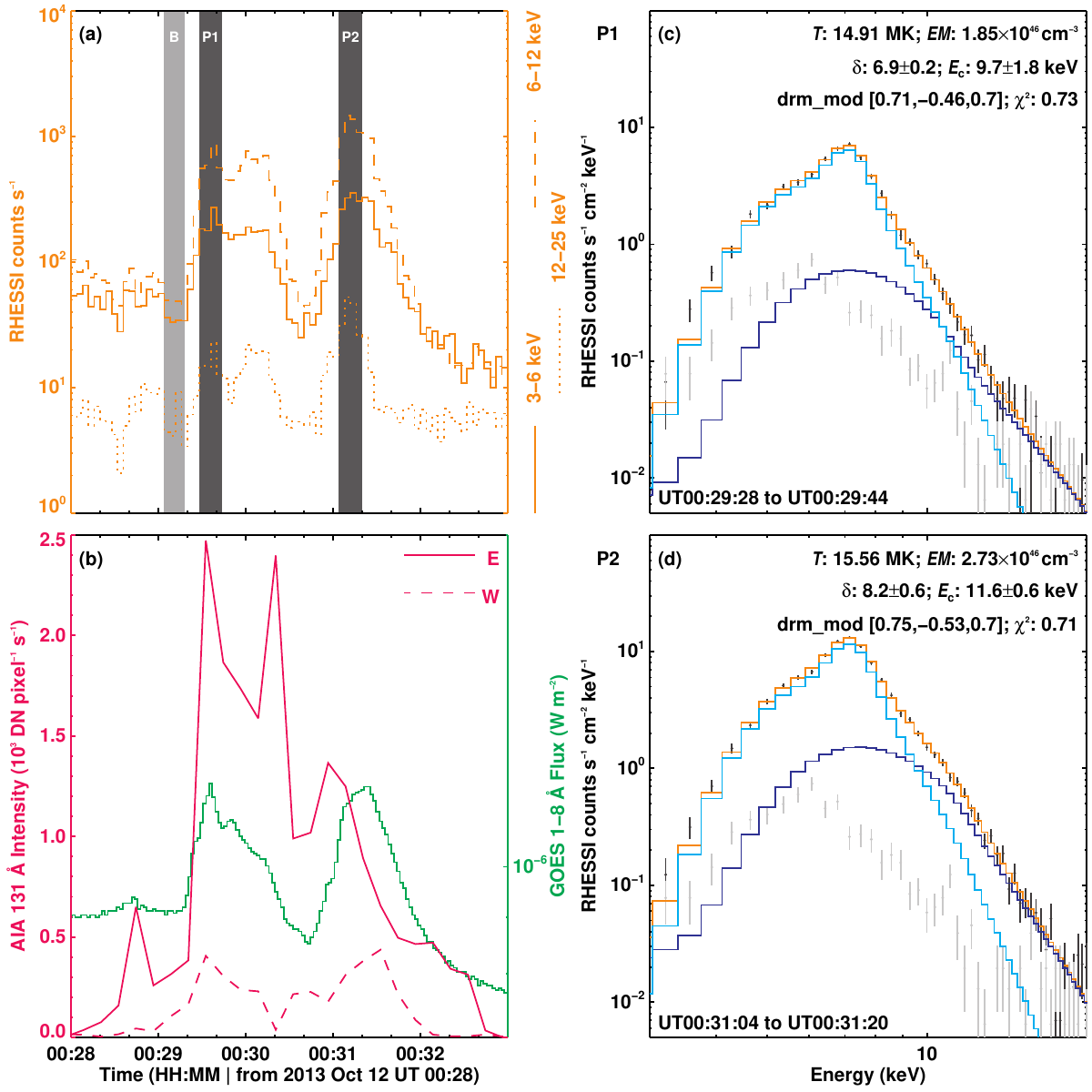}
\caption{Temporal and thermal evolution. In panel (a) we plot the time series of RHESSI HXR count rates in three energy windows. The gray bars identify the background (B) and two peaks (P1 and P2) in the RHESSI time series. The disk-integrated GOES soft X-ray time series (green) and cotemporal AIA EUV 131\,\AA\ filter light curves of footpoints E and W (pink; see Fig.\,\ref{fig:map}i) are plotted in panel (b). Panel (c) shows the RHESSI energy spectrum as dark gray points of peak P1, binned over the time period defined by the width of the corresponding darker gray bar in panel (a). The lighter gray points are the background energy spectrum binned over the time period defined by the lighter gray bar B in panel (a). Fits to the spectrum are shown (cyan: thermal; blue: thick2; orange: total). The resulting temperature ($T$), emission measure ($EM$), electron spectral index ($\delta$), energy cutoff ($E_{\rm c}$), parameters of the pseudo-function (\texttt{drm\_mod}), and goodness of the fit ($\chi^2$) are listed. Panel (d) is same as panel (c), but plotted for the RHESSI peak P2 from panel (a) (see Sects.\,\ref{sec:obs} and \ref{sec:flare} for details).\label{fig:lc}}
\end{center}
\end{figure*}

The main body of the UV brightening captured in the AIA 1600\,\AA\ and in some IRIS wavelength ranges is completely absent in the EUV filter images of AIA. We instead mainly observe the fibrillar material as a darker absorption feature in the EUV (Fig.\,\ref{fig:map}h--i). This darker EUV structure makes a different angle with respect to the Mg\,{\sc ii}\,k core fibrils, however. While the EUV fibrils run diagonally from the northwest to the southeast in the field of view, the Mg\,{\sc ii}\,k core fibrils have a nearly north-south orientation. This might indicate shear in the magnetic field with height at this location. Moreover, we note a pair of highly compact brightenings at the supposed footpoint locations of the cusp-shaped loop feature. While the eastern footpoint E is visible in all the considered EUV passbands of AIA, the western footpoint W is more prominent in the AIA 94\,\AA\ and 131\,\AA\ filters (Fig.\,\ref{fig:map}h--i), which receive contributions to emission from plasma at temperatures in excess of 7\,MK. This suggests that part of the event might be heated to higher coronal temperatures.

Based on our spatial morphological comparison between the UV and the EUV wavelength ranges, we suggest that the observed event is overall similar to that of a UV burst residing below the chromospheric canopy. However, unlike a classical UV burst that occurs on smaller scales of 1\arcsec--2\arcsec\ \citep[e.g.,][]{2014Sci...346C.315P,2018SSRv..214..120Y}, the observed event is much larger, at scales of about 10\arcsec. Furthermore, based on different parts of the burst that were detected in the UV and EUV, we suggest that the event might harbor multithermal plasma, which we discuss in the following section. 

\section{Flare characteristics and high temperatures\label{sec:flare}}

Cotemporal with IRIS and AIA observations of the UV burst, GOES recorded two small peaks in the soft X-ray emission in its 1--8\,\AA\ band around UT\,00:30 that were equivalent to the flux of a C-class flare. Cospatial and cotemporal with the UV burst, RHESSI also detected a compact bright source with energies of up to 25\,keV. The RHESSI source overlaps footpoint W (see the overlaid contours in Fig.\,\ref{fig:map}). Although it is compact, we decomposed the RHESSI source region into a larger region that emitted at lower energies of 5--8\,keV and two smaller footpoint-like regions at the ends of the larger section, which emitted at higher energies ($>11$\,keV). The temporal evolution of the EUV to HXR emission patterns that covered the burst region is plotted in Fig.\,\ref{fig:lc}a--b. The GOES and HRX light curves show two distinct peaks with a potential substructure in the first peak. The AIA 131\,\AA\ EUV light curves are similar. In particular, the emission evolution from footpoint W correlates well with the X-ray light curves. This further confirms the spatial location of the RHESSI source with respect to the IRIS and AIA features. 

\begin{table*}
\setlength{\tabcolsep}{8pt}
\renewcommand{\arraystretch}{1.25}
\begin{center}
\caption{Fit parameters of the two HXR flares from Fig.\,\ref{fig:lc}.\label{tab:flare}}
\begin{tabular}{c c c c c c c}
\hline\hline
Flare & $T$ & $EM$  & $N$ & $\delta$ & $E$$_{\rm c}$ & $NT Pow$ \\
&  [MK] & [$\times10^{46}$\,cm$^{-3}$] &  [$\times10^{35}$\,e$^{-}$\,s$^{-1}$] &  & [keV] & [$\times10^{27}$\,erg\,s$^{-1}$] \\
\hline
P1 & 14.91$\pm$0.53 & 1.85$\pm$0.34 & 0.85$\pm$0.95 & 6.94$\pm$0.15 & 9.74$\pm$1.83 & 1.60$\pm$1.81 \\ 
P2 & 15.56$\pm$0.52 & 2.73$\pm$0.39 & 1.47$\pm$0.34 & 8.21$\pm$0.64 & 11.63$\pm$0.64 & 3.17$\pm$0.77 \\ 
\hline
\end{tabular}
\end{center}
\end{table*}

We used the Object Spectral Executive \citep[OSPEX software;][]{2002SoPh..210..165S} to fit the RHESSI HXR spectra during two different peaks in the X-ray light curves that are shown in in Fig.\,\ref{fig:lc}. These spectra were background-subtracted using the pre-flare time before they were fit with an isothermal model assuming coronal abundances \citep[][]{1992PhyS...46..202F}, combined with a nonthermal model \citep{2011SSRv..159..107H} representing the cold thick target emission from a negative power-law distribution of electrons (with a spectral index $\delta$ above a low-energy cutoff of $E_\mathrm{C}$). In addition, we used a pseudo-function (\texttt{drm\_mod}) that accounted for the degraded RHESSI detector performance (poorer energy resolution and increased noise due to radiation damage) by fitting additional parameters for the spectral full width at half maximum resolution fraction and energy gain offset. The spectral analysis was performed individually for the four RHESSI detectors (1, 4, 6, and 9) with the best spectral performance at the time of the flare. All of these detectors give similar fit results. The fitted spectra for the detector with the best performance (Detector 1 is the least noisy, and has fit parameters with the smallest uncertainties; Table\,\ref{tab:flare}) are shown in Fig.\,\ref{fig:lc}c--d. The fitted temperature, $T$, is about 15\,MK during flare P1. It slightly rises to about 16\,MK during flare P2. The emission measure, $EM$, also slightly increased during flare P2. The nonthermal emission during flare P1 is more marginal (the uncertainties of the total electron flux $N$ and of the resulting total nonthermal power, $NT Pow$, are consistent with null emission due to the noise in the spectrum at energies $>10$\,keV; see Fig.\,\ref{fig:lc}c), but clearer during flare P2, with a nonthermal power of about $3\times10^{27}$\,erg\,s$^{-1}$ (see Fig.\,\ref{fig:lc}d). Our observations thus provide evidence for particle acceleration in this flare.

The EUV emission patterns indicate the existence of multithermal plasma. Footpoint E shows concurrent brightening in all the filters. Stronger enhancements are visible in the 171\,\AA, 193\,\AA, and 211\,\AA\ filter images. Even in the AIA 304\,\AA\ passband, which samples the lower transition region plasma around 50\,kK, we therefore observed features that are similar to those seen in the 193\,\AA\ filter. This implies that the region primarily comprises cooler plasma, but also a possibly hotter component. In contrast, footpoint W is more prominently visible mainly in the 94\,\AA\ and 131\,\AA\ filter images, and the enhancements in the 171\,\AA, 193\,\AA, and 211\,\AA\ images are relatively weaker. This further suggests that emission from the hotter component must dominate the signal in the 94\,\AA\ passband, as is also likely the case for the 131\,\AA\ emission. This is also consistent with the spatial overlap of the RHESSI source with footpoint W in the AIA images. 

The UV burst definition given by \citet{2018SSRv..214..120Y} specifically excluded flares. Because of the clear UV-burst-like characteristics and the simultaneous HXR emission exhibited by the brightening, we propose that the observed event is a hybrid phenomenon that combines a UV-burst and a flare. The UV spectrum generally contains the definitive diagnostic information for an event like this.

\begin{figure*}
\begin{center}
\includegraphics[width=\textwidth]{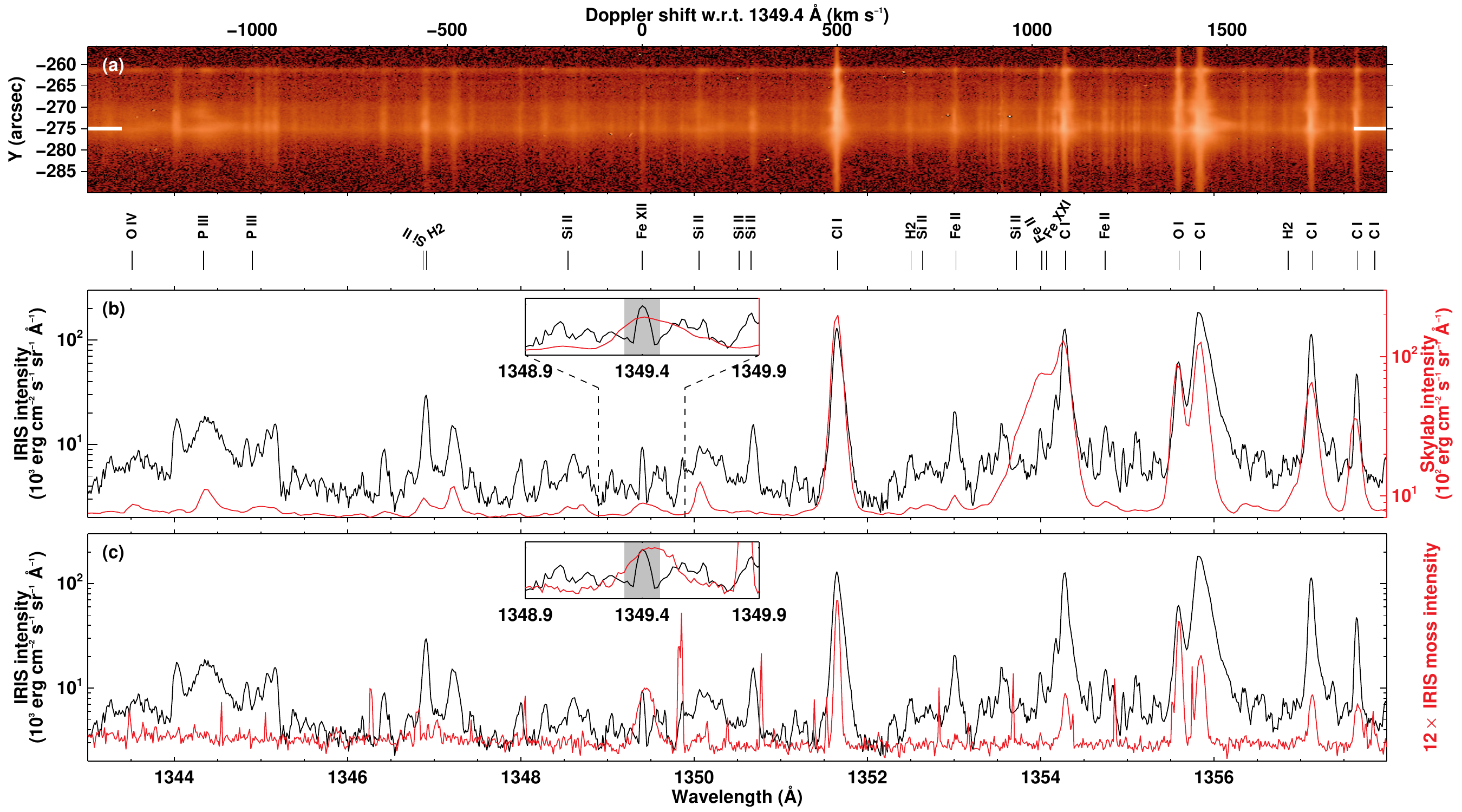}
\caption{Unusual spectral characteristics. Panel (a) displays the IRIS detector image. The intensity is plotted on a log-scale along the solar Y-axis that cuts through the cyan diamond in Fig.\,\ref{fig:map}g. The black spectrum in panels b--c is the IRIS intensity obtained from the pixel marked by a pair of white whiskers in the upper panel (the same as the location of the cyan diamond pixel in Fig.\,\ref{fig:map}g). The red spectrum in panel b is the intensity of a flare observed by Skylab. We applied a constant $+70$\,m\AA\ offset to the original Skylab wavelength array to match the positions of prominent IRIS spectral lines in the plotted wavelength range. In the inset, we zoom into the wavelengths around 1349.4\,\AA\ (nominal rest wavelength of Fe\,{\sc xii}) and compare the spectra from the IRIS hybrid event and the Skylab flare. The width of the gray shaded bar represents the expected thermal full width at half maximum of the Fe\,{\sc xii} emission line (0.152\,\AA; including the IRIS instrumental width of 26\,m\AA). The red curve in panel c is an average spectrum from a moss region (multiplied by 12) that was observed by IRIS in the same data set. In the inset, we compare the spectra from the IRIS hybrid event and the moss region around 1349.4\,\AA. The labels of some of the known spectral species in the plotted wavelength range are added for reference. The line list was retrieved from the \texttt{iris\_chianti\_lookup\_table} that is available in the IRIS database and \citet{1986ApJS...61..801S} (see Sects.\,\ref{sec:obs} and \ref{sec:spec} for details).\label{fig:cdet1}}
\end{center}
\end{figure*}

\section{Far-UV spectral characteristics\label{sec:spec}}

In addition to the distinct UV and EUV spatial morphological features, the observed hybrid event also displayed intriguing spectral features, in particular, in the FUV range. We examined the spectral profiles that emerged from two locations within the event in the wavelength range 1343\,\AA\ to 1358\,\AA, which includes the Fe\,{\sc xii} 1349.4\,\AA\ and the Fe\,{\sc xxi} 1354\,\AA\ spectral lines.

The UV emission from the cusp region is rich. It includes several spectral lines, and neutral lines exhibiting high intensities (Fig.\,\ref{fig:cdet1}a). Although at least some burst material moved along the spine from the cusp region, the neutrals did not show any large line-of-sight Doppler motions with respect to their nominal line positions at rest (middle panel of Fig.\,\ref{fig:cdet1}). Similarly, the Ni\,{\sc ii} spectral line around 1393.33\,\AA, which is superimposed on the Si\,{\sc iv} line (Fig.\,\ref{fig:map}f), is not clearly Doppler shifted with respect to its nominal rest wavelength. This suggests some decoupling between the ambient atmosphere and the burst-flare hybrid phenomenon.

We also compared the IRIS spectral profiles with those from an X-class flare observed by Skylab (Fig.\,\ref{fig:cdet1}b). Despite differences in the flare class (and thus, in the associated magnetic energy content), these two spectral profiles match remarkably well. However, the Skylab flare spectrum reveals clear signatures of the Fe\,{\sc xii} 1349.4\,\AA\ and the Fe\,{\sc xxi} 1354\,\AA\ spectral lines, which are not obvious in the IRIS data. To demonstrate these differences further, we show an enlarged Fe\,{\sc xii} 1349.4\,\AA\ line region in the inset of Fig.\,\ref{fig:cdet1}b. While the IRIS data show a spectral line, its thermal width is even smaller than the expected Fe\,{\sc xii} line width. Skylab recorded a width of the Fe\,{\sc xii} line that exceeded its thermal width, however.

To confirm whether the lack of Fe\,{\sc xii} 1349.4\,\AA\ is an artifact in this particular data set, we compared the burst profile with spectral lines that emerged from a nearby moss region in the same data set (Fig.\,\ref{fig:cdet1}c). A clear Fe\,{\sc xii} signal in the moss region is visible. This confirms the lack of Fe\,{\sc xii} 1349.4\,\AA\ and the Fe\,{\sc xxi} 1354\,\AA\ spectral lines in the observed cusp region.

We then compared the IRIS lines profiles from the western edge of the burst feature close the RHESSI source region (see the cyan circle in Fig.\,\ref{fig:map}). These spectra were obtained around UT\,00:35, some 4\,minutes after the peak of the RHESSI flare. During this late phase, EUV to HXR emission was nearly absent at that location, but IRIS continued to show enhanced line profiles (Fig.\,\ref{fig:cdet2}). Conclusive evidence for the existence of Fe\,{\sc xxi} 1354\,\AA\ spectral lines is lacking here as well. The enhanced and broad line profiles mean in general that the lower atmosphere is continually heated by magnetic energy release and deposition in the lower atmosphere well after the flare peak passed, however.

\begin{figure}
\begin{center}
\includegraphics[width=0.49\textwidth]{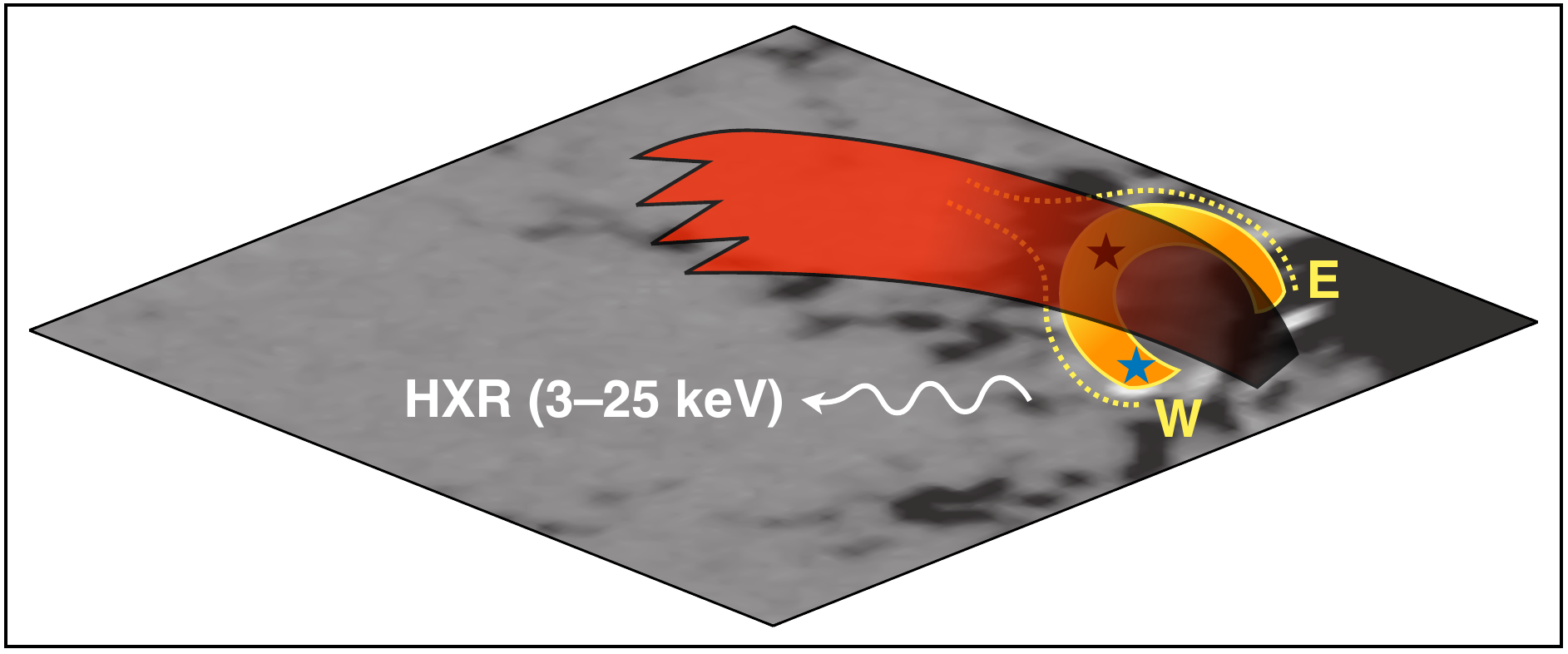}
\caption{Possible scenarios for the hybrid phenomenon. The background grayscale image is the observed HMI magnetogram (same as in Fig.\,\ref{fig:map}b), the yellow shaded region is the UV burst loop feature rooted at footpoints W and E, and the red shaded structure is the overlying chromospheric canopy. RHESSI detected HXR emission in the 3-25\,keV range at footpoint W (see Fig.\,\ref{fig:map}). The stars indicate the location of the magnetic energy release (blue: energy release at footpoint W; black: energy release at the loop-top cusp; see Sect.\,\ref{sec:disc} for a discussion).\label{fig:illus}}
\end{center}
\end{figure}

\section{Discussion\label{sec:disc}}
The main body of the brightening is not apparent in the IRIS Mg\,{\sc ii}\,k core map. Moreover, except for the two compact footpoints, the AIA EUV images show a darker absorption feature at the location of the burst. These two aspects emphasize the point that the event was located in the lower atmosphere. We propose two possible scenarios to explain the observed hybrid phenomenon. In the first scenario, magnetic energy that is released at the cusp-shaped loop top below the chromospheric canopy (black star in Fig.\,\ref{fig:illus}) triggers the UV burst and is deposited at both footpoints. This deposited energy is the source of the HXR emission at footpoint W. This is similar to the scenario of coronal loop-top magnetic reconnection in flares \citep[][]{1994Natur.371..495M}, but now at the top of a loop that is embedded in the chromosphere. In the second scenario, energy that is directly released at footpoint W causes the HXR emission at this location (blue star in Fig.\,\ref{fig:illus}). Part of the energy is also transported from footpoint W to footpoint E. In this case, magnetic energy that is released in the lower atmosphere (close to the photospheric footpoints), not at coronal heights, drives the observed flare dynamics. Neither case requires the site of energy release to be located in the corona. The event that we studied had remarkably brief GOES emissions, which are intuitively consistent with compact spatial structures and high plasma densities. 

The two scenarios that indicate single reconnection sites are rather simplistic. The surface magnetic field distribution underlying the event is quite complex, with scattered mixed-polarity fields (Fig.\,\ref{fig:map}b). Therefore, it is highly plausible that multiple reconnection sites were situated at different heights above the surface and caused the observed event. In particular, the brightening at footpoint E and the HXR emission at footpoint W could have been triggered by different reconnection events at different heights. The near concurrence of the brightenings at these footpoints, along with the GOES and HXR signals, suggests a coupling of the reconnection events, however. The highly inclined penumbral magnetic field, coupled with the complex mixed-polarity magnetic field distribution, might have caused the heating of the plasma to over 10\,MK, as indicated by the HXR emission, below the chromospheric canopy.

The fitting of HXR spectrum also shows evidence of nonthermal particles, at least in flare P2 (Fig.\,\ref{fig:lc}, Table\,\ref{tab:flare}). Alfv{\'e}n wave pulses \citep[][]{2008ApJ...675.1645F} combined with the effects of local reacceleration in the HXR source region \citep[][]{2009A&A...508..993B} could accelerate particles in the otherwise collision-dominated chromosphere \citep[][]{2010MmSAI..81..616F}. Furthermore, the prolonged heating we inferred from the enhanced FUV spectral lines in IRIS data suggests that the energy conversion might be related to magnetic reconnection and turbulence in the lower atmosphere, which is invoked to be the case in the early stages of flares heating \citep[e.g.,][]{2013ApJ...774..122H,2018SciA....4.2794J,2020ApJ...890L...2C}. 

Many flares are shown to exhibit the so-called hot X-ray onset of plasma temperatures in the range of 10--15\,MK, which is linked to the brightening of low-lying loop regions \citep[][]{2021MNRAS.501.1273H,2023A&A...679A.139B}. In our observations, the energy release and conversion evidently occurs below the chromospheric canopy. This indicates a scenario in which the hot onsets in low-lying loops are related to magnetic heating within the chromosphere. Thermal conduction or Alfv{\'e}n waves in the turbulent environment play an important role in the energy transport there because nonthermal sources in the corona, if any, would only appear at the start of the main energy release phase \citep[][]{2009A&A...498..891B}.

Complex magnetic fields and flux cancellation in the photosphere underlie the observed flare. This implies that magnetic reconnection is the driver of the hybrid phenomenon (Fig.\,\ref{fig:map}). Signatures of magnetic reconnection associated with mixed-polarity magnetic field interactions are observed at the feet of hot loops in active region cores \citep[][]{2018A&A...615L...9C,2020A&A...644A.130C}. As mentioned in Sect.\,\ref{sec:obs}, the observed hybrid phenomenon is also similar to violent active region multithermal jets \citep[][]{2008A&A...491..279C}, which often involve eruption of mini-filaments that form in the sheared magnetic arcades \citep[e.g.,][]{2016ApJ...821..100S,2017ApJ...844...28S,2021A&A...645A..80J}, but buried in the chromosphere. Our event is also similar to the flaring arch filament observed by \citet{2020ApJ...893L..13S}. The authors found signatures of absorption lines superimposed on transition region emission lines in a small white-light flare and suggested reconnection in the lower atmosphere as a trigger of the event. Thus, flux emergence and cancellation at the surface causing lower atmospheric reconnection would then be a unified model to explain compact impulsive heating events in the solar atmosphere from the chromosphere through the corona \citep[][]{2018ApJ...862L..24P}. 

Current sheets, in which magnetic energy is converted into other forms, are central to magnetic reconnection \citep[][]{2000mare.book.....P,2002A&ARv..10..313P,2022LRSP...19....1P}. The formation of thin, elongated current-sheet-like features in flares also marks the phase of enhanced high-energy HXR emission in some flares \citep[][]{2021ApJ...911..133C}. Our observations lacked clear evidence of these current-sheet resembling plasma features. This either means that these features cannot be resolved in current observations or that reconnection occurs concurrently throughout the 3D volume of the loop \citep[][]{2022LRSP...19....1P}.

The Fe\,{\sc xii} 1349.4\,\AA\ spectral line that IRIS could observe in principle is generally weak and arises from a forbidden transition. Only a few studies have discussed observations of this line based on IRIS observations so far \citep[e.g.,][]{2016ApJ...827...99T,2019ApJ...871...82G,2022ApJ...930...61D}. Despite clear signatures of hot plasma in our observations, the lack of Fe\,{\sc xii} 1349.4\,\AA\ and the Fe\,{\sc xxi} 1354\,\AA\ spectral lines is puzzling (Figs.\,\ref{fig:lc}, \ref{fig:cdet1}, \ref{fig:cdet2}). The narrower line caught by IRIS at the location of the nominal Fe\,{\sc xii} 1349.4\,\AA\ line could be an unknown blended line, as was recently pointed out \citep[][]{2022ApJ...930...61D}. The question why Skylab observed this line clearly but IRIS did not remains open in this case. It might perhaps be related to the lack of coronal energy release in our hybrid event compared to the X-class flare observed by Skylab.

Events with intriguing emission characteristics are common in the solar atmosphere. White-light flares, for example, may obviously have a photospheric origin, although the emission physics currently remains uncertain. In one event, \citet{2012ApJ...753L..26M} directly estimated the geometric height of white-light and HXR emissions. They located them close to the $\tau_{5000}=1$ definition of the photosphere in terms of the optical depth at 5000\,\AA.

The entire structure of a flare may develop on spatial scales that are so compact as to imply that they do not geometrically extend into the corona. For instance, \citet{1992ApJ...384..656W} reported an impulsive solar radio burst that originated from a highly dense (electron density $\sim5\times10^{11}$\,cm$^{-3}$) single-loop structure in the solar corona. The event exhibited a flat spectrum in the range of 15 to 86\,GHz, with a sharp cutoff in the spectrum between 5.0 and 8.6\,GHz. While the flat spectrum might indicate optically thin thermal bremsstrahlung emission, the authors argued against this emission mechanism because they did not observe clear soft X-ray emission from the event. The study presented the case for optically thick emission due to thermal absorption of nonthermal gyrosynchrotron emission, or optically thin gyrosynchrotron emission resulting from absorption by high-density material in the line of sight. These events, including the event we discussed here, may indicate gaps in our understanding of the multithermal coupling of the solar atmosphere during impulsive heating events.

\section{Conclusions}

Using multiwavelength observations ranging from the near-UV to hard X-rays, we identified a hybrid explosive phenomenon in the solar atmosphere with properties of UV bursts and also of flares \citep[see also][]{2020ApJ...893L..13S}. The event was confined to altitudes below the chromospheric canopy, suggesting that the magnetic energy that heated the plasma and accelerated the particles was released in the lower atmosphere. Enhanced spectral profiles that sampled the lower atmosphere were observed even minutes after the flare peak in both soft and hard X-rays, suggesting a prolonged release of magnetic energy and subsequent plasma heating. The IRIS FUV wavelength range is quite rich, but we found remarkable similarities between the spectral profiles of the C-class flare we studied and the Skylab observations of an X-class flare. This indicates common processes of energy conversion in the lower atmosphere in distinct classes of flares. It might prove to be challenging, but efforts to model this hybrid phenomenon for its release and transport of energy along with its radiative properties will shed light on the intricacies of the solar atmospheric coupling and its common behavior from small UV bursts to major flares \citep[][]{2019A&A...627A.101V,2024MNRAS.527.2523K}.

\begin{acknowledgements}
We thank the referee for helpful comments that improved the presentation of the manuscript. We are grateful to Paulo J.~A. Sim{\~o}es for making the calibrated Skylab data available. L.P.C. thanks Smitha Narayanamurthy (MPS) for useful discussion on spectral line identification. L.P.C. gratefully acknowledges funding by the European Union (ERC, ORIGIN, 101039844). Views and opinions expressed are however those of the author(s) only and do not necessarily reflect those of the European Union or the European Research Council. Neither the European Union nor the granting authority can be held responsible for them. AIA and HMI are instruments on board the Solar Dynamics Observatory, a mission for NASA's Living With a Star program. IRIS is a NASA small explorer mission developed and operated by LMSAL with mission operations executed at NASA Ames Research Center and major contributions to downlink communications funded by ESA and the Norwegian Space Centre. We are grateful to RHESSI and GOES teams for making the data publicly available. This research has made use of NASA’s Astrophysics Data System. We acknowledge the usage of {\sc AIAPY} open source software package \citep[v0.7.4;][]{Barnes2020,Barnes2021}. CHIANTI is a collaborative project involving George Mason University, the University of Michigan (USA), University of Cambridge (UK) and NASA Goddard Space Flight Center (USA). This research was supported by the International Space Science Institute (ISSI) in Bern, through ISSI International Team project \#23-586 (Novel Insights Into Bursts, Bombs, and Brightenings in the Solar Atmosphere from Solar Orbiter).
\end{acknowledgements}

\begin{appendix}
\section{Spectral line comparisons\label{sec:app}}
\begin{figure*}
\begin{center}
\includegraphics[width=\textwidth]{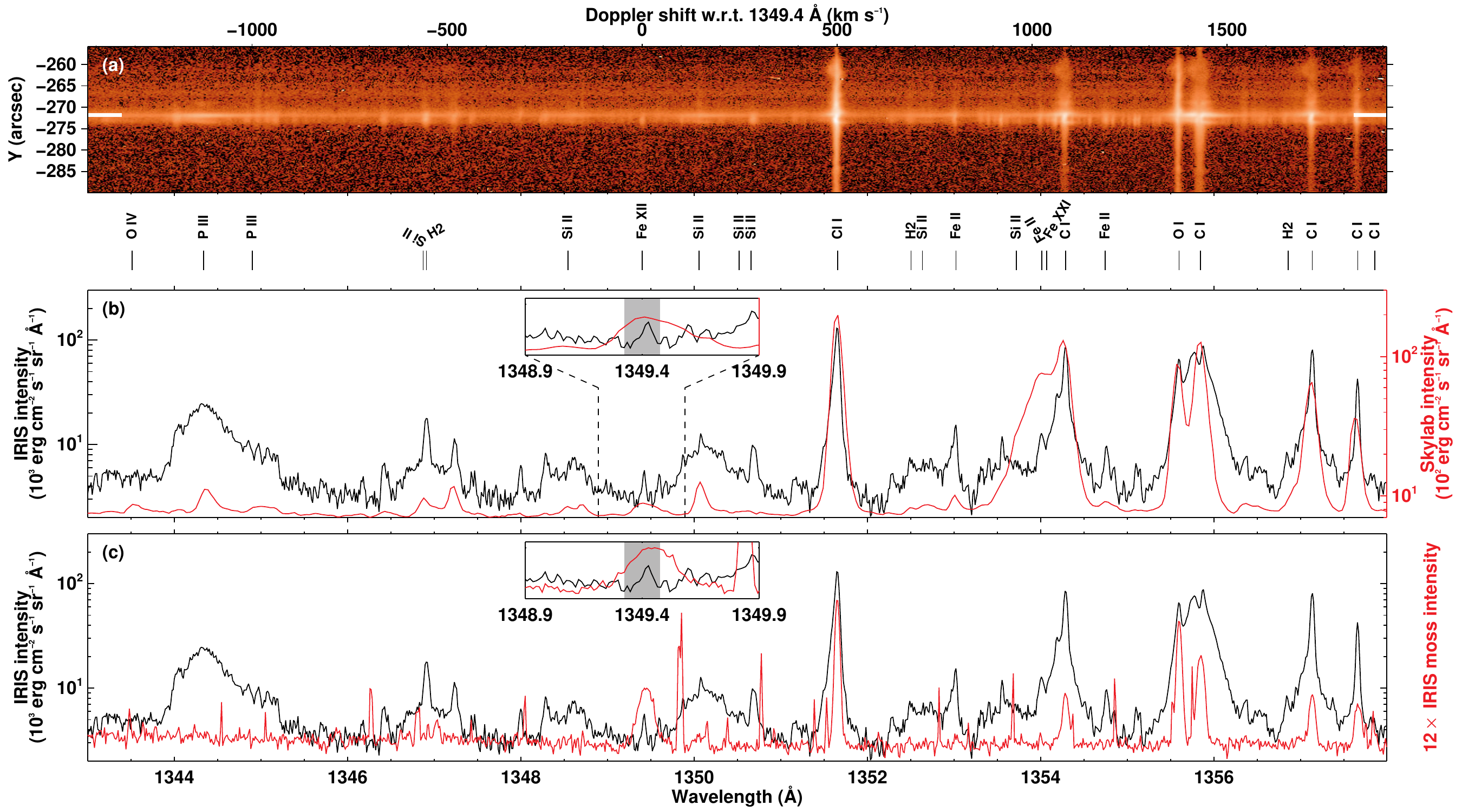}
\caption{Spectral line comparisons. Same as Fig.\,\ref{fig:cdet1} but plotted along the solar Y-axis cutting through the cyan circle in Fig.\,\ref{fig:map}g. In panels b--c the red curves remain the same as in Fig.\,\ref{fig:cdet1}. See Sects.\,\ref{sec:obs} and \ref{sec:spec} for details.\label{fig:cdet2}}
\end{center}
\end{figure*}

\begin{figure*}
\begin{center}
\includegraphics[width=\textwidth]{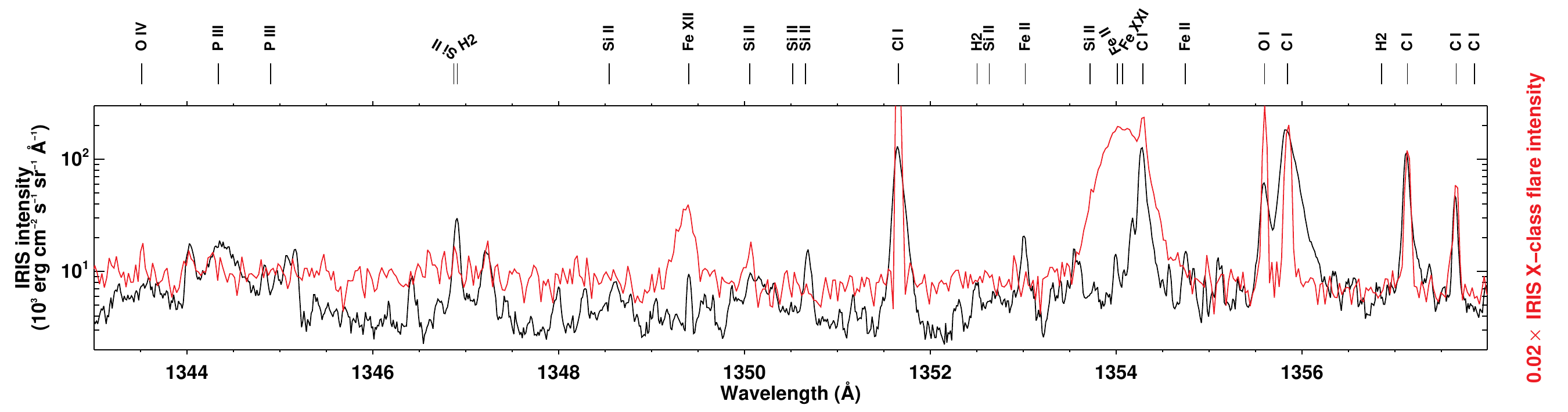}
\caption{Spectral line comparisons. Same as Fig.\,\ref{fig:cdet1}b but the red curve here corresponds to IRIS spectral lines from a flare ribbon of an X-class flare on 2014 October 25 (IRIS pixel coordinate $=222$). See Sects.\,\ref{sec:obs}, \ref{sec:spec} and Appendix\,\ref{sec:app} for details.\label{fig:cdet3}}
\end{center}
\end{figure*}

We present additional examples of unusual spectral characteristics of the hybrid event. We discussed Fig.\,\ref{fig:cdet2} at the end of Sect\,\ref{sec:spec}. In Fig.\,\ref{fig:cdet3} we compare the spectral profile from the hybrid event with an X-class flare that IRIS observed on 2014 October 25. According to GOES observations, the flare peak a was around UT\,17:05. The observational details of the X-class flare are as follows. The IRIS slit was scanning the core of the active region NOAA AR12192 in a sit-and-stare mode between UT\,14:58:28 and UT\,18:00:36 (OBS ID: 3880106953). These data have a spatial scale of about 0.3\arcsec\ along the slit direction. IRIS acquired a full readout of the detector. The spectra were recorded with an exposure time of 4\,s. We radiometrically calibrated the level-2 data. This particular dataset was used by \citet{2019ApJ...878..135K} to study the heating properties in umbral flare kernels.

The red colored profile in Fig.\,\ref{fig:cdet3} is from the site of a flare ribbon, obtained by averaging the data during the phase between UT\,17:24 and UT\,17:30, when the GOES X-ray flux was still close to its peak value in the 1-8\,\AA\ band. Both the Fe\,{\sc xii} and Fe\,{\sc xxi} spectral line features are clear in this X-class flare, which is not the case with the hybrid event (black curve in Fig.\,\ref{fig:cdet3}).

\end{appendix}

\end{document}